# Earth Virtualization Engines - A Technical Perspective

**Torsten Hoefler**, ETH Zurich, Zurich, 8092, Switzerland,
**Bjorn Stevens**, Max Planck Institute for Meteorology, Hamburg, 20146 Germany,
**Andreas F. Prein**, National Center for Atmospheric Research, Boulder, 80301, CO, USA,
**Johanna Baehr**, Center for Earth System Research and Sustainability, Universität Hamburg (CEN), 20146 Germany,
**Thomas Schulthess**, ETH Zurich and Swiss National Supercomputing Center (CSCS), Lugano, 6900, Switzerland,
**Thomas F. Stocker**, Climate and Environmental Physics, and Oeschger Centre, University of Bern, 3012, Switzerland,
**John Taylor**, Commonwealth Scientific Industrial Research Organisation, Canberra, ACT 2601, Australia,
**Daniel Klocke**, Max Planck Institute for Meteorology, Hamburg, 20146 Germany,
**Pekka Manninen**, CSC - IT Center for Science, Espoo, 02101 Finland
**Piers M. Forster**, University of Leeds, Leeds LS2 9JT, UK,
**Tobias Kölling**, Max Planck Institute for Meteorology, Hamburg, 20146 Germany,
**Nicolas Gruber**, ETH Zurich, Zurich, 8092, Switzerland,
**Hartwig Anzt**, University of Tennessee, Knoxville, 37996, USA, Germany,
**Claudia Frauen**, German Climate Computing Center (DKRZ), 20146 Hamburg, Germany,
**Florian Ziemen**, German Climate Computing Center (DKRZ), 20146 Hamburg, Germany,
**Milan Klöwer**, Massachusetts Institute of Technology, Cambridge MA 02139, USA,
**Karthik Kashinath**, NVIDIA Corporation, Santa Clara, California 95051, USA,
**Christoph Schär**, Atmospheric and Climate Science, ETH Zurich, 8092 Zürich, Switzerland,
**Oliver Fuhrer**, Federal Office of Meteorology and Climatology MeteoSwiss, Zurich, 8058, Switzerland,
**Bryan N. Lawrence**, NCAS & Departments of Meteorology and Computer Science, University of Reading, Reading, UK.

*Participants of the Berlin Summit on Earth Virtualization Engines (EVEs) discussed ideas and concepts to improve our ability to cope with climate change. EVEs aim to provide interactive and accessible climate simulations and data for a wide range of users. They combine high-resolution physics-based models with machine learning techniques to improve the fidelity, efficiency, and interpretability of climate projections. At their core, EVEs offer a federated data layer that enables simple and fast access to exabyte-sized climate data through simple interfaces. In this article, we summarize the technical challenges and opportunities for developing EVEs, and argue that they are essential for addressing the consequences of climate change.*

We are all witnessing the effects of climate change. Hotter summers, prolonged droughts, massive flooding, or ocean heat waves are examples of extreme weather and climate events that are growing in frequency and intensity. Many agree that addressing climate mitigation and adaptation is the biggest problem humanity faces today. *A large group of scientists and practitioners from different climate-related domains, including some computer scientists, got together for a week in Berlin this July to discuss the concept of "Earth Virtualization Engines" (EVEs).* The summit kicked off with the question: *"If climate change is the most critical problem today, why are we not using the largest computers to help solve it?"*.

The question proved provocative in two ways. One, many people think we are using them already, although we aren't, while others fear that a focus on technologies masks the many other, mostly human and social dimensions of the problem. Through intense discussions over multiple days, the participants were able to reconcile these different viewpoints in their summary statement. Here, a subset of the participants outline their common understanding of the technical landscape and constraints and a possible technical realization. *This article distills many productive conversations into a set of observations and ideas to form a basis to guide future investments and more detailed investigations*. This is by no means meant as a final design, it should be seen as an initial technical contribution to the journey that will enable humanity to assess the detailed risks of climate change from the local to the global scale.

*Climate models are currently our central source of information on how climate will continue to change in the future*. Thus, any improvement we make to their fidelity, combined with improved theoretical understanding, is likely to directly benefit the quality of climate prediction. Today's exascale computing will enable global climate simulations with a grid resolution approaching 1 kilometer within the next few years. This resolution allows us to explicitly resolve a number of critical physical processes, such as convection in the atmosphere, which are not captured correctly in models typically employed today. Some of the largest improvements when using high resolution models are seen in the simulation of key variables such as precipitation or in capturing orographic effects [prein15]. However, such km-scale models produce tens of exabytes of data to analyze for diverse uses. Additionally, these *climate models are computationally expensive, hard to optimize, let alone use, and link to climate impact studies*. This greatly limits the number of groups worldwide who can operate these models, contribute to their development and benefit from the data they generate.

The Berlin EVE summit recognized that in order to cope with the consequences of climate change, many more users have to have equal access to such climate data and the capability to extract information relevant to them. To this end, one of the most critical elements of EVE consists of an *interactive data access layer that allows simple navigation, extraction, and application of the climate simulation data*. In a simple form, we imagine an interface such as Google's Earth Engine, or Microsoft's

Planetary Computer's "Explore" interface (see Fig. 1) – albeit with much better climate predictions. Behind the scenes, the data of these predictions will be served by a distributed system federating multiple centers running kilometer-scale models as data generators and distilled into actionable information by a layer of custom-built climate tools and services. Such an interactive data layer would also empower local communities – agronomists in India, water managers in Peru – to link their tools to the data. Furthermore, accessing details of the dataset, climate scientists can better understand and reason about simulated changes. Deploying and operating such an interactive system is a grand infrastructure and engineering challenge. Given its intended use, care must be exercised to ensure that its use is not restricted to industry (e.g., cloud providers such as Microsoft) or large-scale research laboratories at the forefront of climate research endeavors (e.g., Max Planck or US DOE).

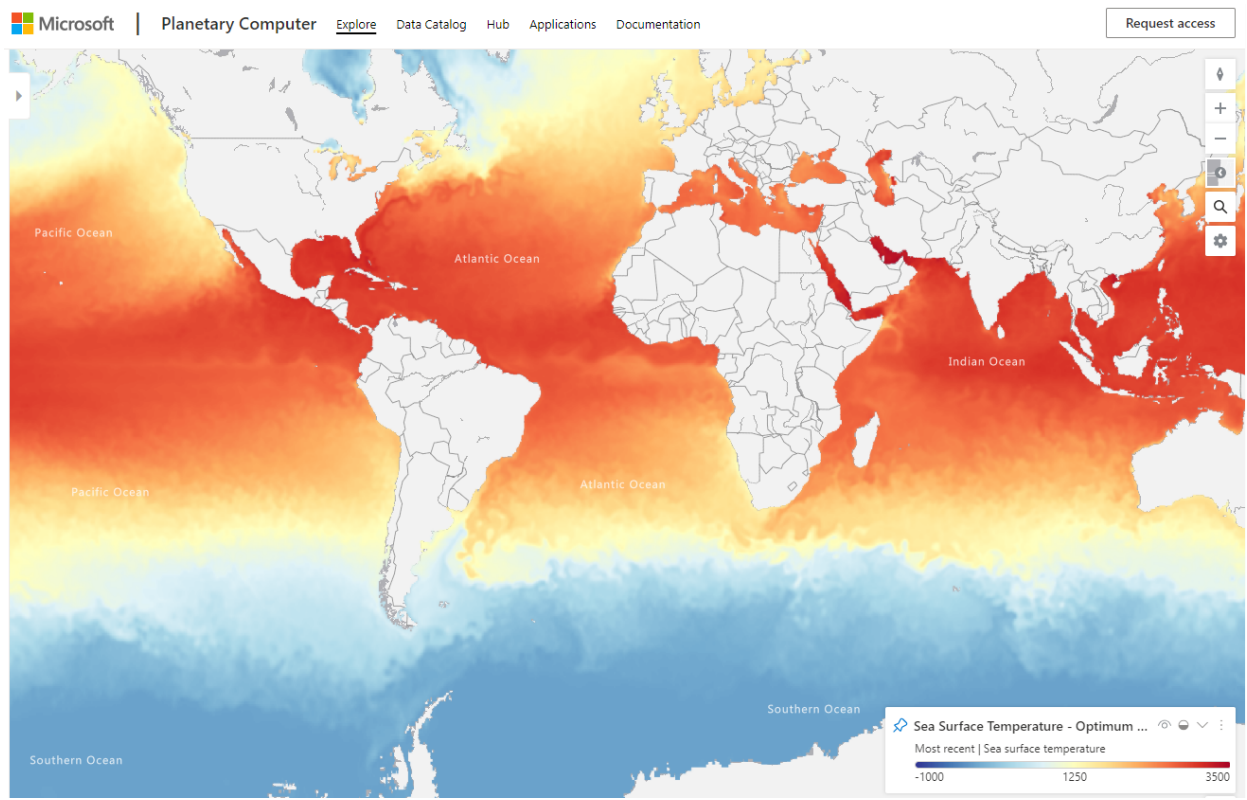

Fig 1: Microsoft Planetary Computer Explorer: Sea Surface Temperature (source: planetarycomputer.microsoft.com).

Machine Learning (ML) techniques, which boosted Artificial Intelligence (AI) with Large Language Models (LLMs), and related techniques are invigorating climate sciences, as they are many other fields. Summit participants recognized the need to combine physics-based simulations with ML techniques to accelerate and improve predictions, and the scope of data-driven methods was widely discussed. Crucially, **ML/AI was identified as essential to opening new opportunities for making the data accessible** by *extracting information that can empower users to act.*

## What is Climate Modeling?

Climate is the statistics of weather typically taken over a 30-year period. Weather is what we observe today and what we can predict in the next few days with relatively small uncertainty given accurate initial conditions. Yet, processes in the atmosphere and ocean are chaotic, which limits our ability to predict the weather beyond two to three weeks ([lorenz72]). Thus, climate modelers do not aim to predict the weather on a specific day in the distant future but rather want to understand how the statistics of weather are changing given alterations in greenhouse gas forcings, aerosols, or other earth system components. Put differently: what type of weather might we expect in a warmer world?

Predicting the coming weekend's weather can be seen as using physics to extrapolate from today's weather. However, predicting the climate in 30 years, with a potentially doubled $CO_2$ concentration requires improved understanding of climate processes. The probability of different weather states may change gradually, but tipping points would accelerate changes, with locally and globally potentially devastating consequences. Understanding, simulating, and especially predicting tipping points is very challenging and Monte-Carlo methods are essential to estimate associated uncertainties. Because it is impossible to know exactly what greenhouse gasses will be emitted in the future and how humans might respond to the impacts of climate change, simulations span a range of different scenarios.

Another difference from a scientific perspective is how we develop and test tools for climate modeling: We can assess the quality of a weather prediction after some days comparing it against reality. However, we only have one realization of past climate change that models are typically tuned to capture and we cannot wait 30 years to understand if our projections were correct. Validating climate projections is thus especially tricky, if not impossible, making proper verification of model implementation essential. Thus, it is extremely important to develop our theoretical understanding and at the same time tightly control statistics and uncertainty!

How do we Model the Climate?

Climate simulations are weather predictions run over several decades to centuries but add physical processes relevant for climate projections to explicitly simulate changing CO2 concentration. The same governing equations are used for atmospheric processes and one can share large parts of the code between weather predictions and climate simulations. The principal difference is that coupled atmosphere-ocean simulations are necessary to simulate climate, and full "Earth System Models" also include land, ice and biosphere models. Most of the computational effort is spent in the atmosphere, although this depends on the complexity of other processes included, such as ice-sheet and vegetation components.

Many weather forecasting centers participated at the EVE Summit, e.g., ECMWF, who offer forecasts as a service. They do this by constantly integrating (or ``assimilating'' [klinker00]) measurement data from satellites, airplanes, or weather stations into a running simulation as an initial condition for a prediction. ECMWF also uses its data assimilation and modeling capabilities to generate reanalysis products that provide best estimates of the historic weather for the past decades [hersbach20].

It's All About the Scales!

Climate simulations discretize the Earth's atmosphere, ocean and land surface into a finite set of roughly equally sized tiles, or voxels, which then determines the simulation's resolution. Each grid box represents an average of the physical or chemical properties within. Today's grid resolutions are about 100 km - i.e., we average an area of approximately 10,000 km² (or about a quarter of Switzerland) into a single number. This makes it impossible to represent processes that are locally relevant, e.g., because they impact parts of cities. Weather modelers have long ago learned the value of simulating at finer resolution, which is, for instance, why the Swiss National Weather Service is developing weather forecast models with a resolution of kilometers. Coarser grid resolutions cause significant uncertainties in climate simulations. Even global metrics such as the Equilibrium Climate Sensitivity (ECS) show high uncertainties at current resolutions. ECS assesses the average increase in Earth's surface temperature, assuming the $CO_2$ in the atmosphere doubled, which ranges from 2-5 ˚C, where the uncertainties are mainly due to the representation of cloud changes [forster21].

The grid resolution indeed has an even bigger impact: physical processes at scales smaller than the grid resolution are not captured by the physical simulation model and need to be represented empirically (using so-called parametrizations). Such parametrizations introduce significant inaccuracies. The accuracy of weather models can be increased significantly by reducing the grid spacing to a few kilometers [schaer21]. Specifically, at such resolutions, the dynamics in deep convective processes, such as large tropical thunderstorms, are simulated explicitly, leading to much higher accuracy in cloud feedbacks [schneider17]. A 100m grid would further improve the representation of shallow convective clouds, another significant step towards higher-fidelity modeling based on first principles, and one particularly relevant for quantities like ECS. Thus, increasing the resolution of climate models is also expected to increase the global prediction quality significantly.

High-resolution climate models are likely our best path forward to improve predictions of the future climate. Yet, a full Earth simulation at kilometer resolution would require about 510 million tiles for each vertical level and second-scale time stepping over decades of simulated time, making simulations exorbitantly expensive.

## Simulation Computational Requirements

Running and calibrating high-resolution climate simulations requires the world's best experts working with the world's best research infrastructures. At the same time, much of the insight in climate research is generated in thousands of small-scale research labs by students and research staff who often do not have easy direct access to the simulation setup and data output. Connecting the simulations to the users through frictionless data access interfaces would empower a wide group of researchers and practitioners, help develop and engage talent globally, and fuel new breakthrough ideas.

Furthermore, the relation between increasing the number of ensemble members to capture the probability distribution better and increased resolution to improve the accuracy of each "trajectory" remains complex. Fig. 2 shows a sketch of this relationship and several axes of potential improvement. We note that the performance measured in simulated years per day (SYPD) scales linearly in the number of ensemble members but worse in resolution. Furthermore, the amount of simulation data produced scales rapidly with the resolution. Code optimization or computing speed improvements of the machine will move us to the right in the plot, towards EVE's target. Machine learning "inside" the simulation to replace costly model components offers a potential joker and can lead to exceeding the goal. ML can also combine the benefits of ensembles and resolution in post-processing "on top" of the simulation to get a better approximation of the distribution [groenquist21].

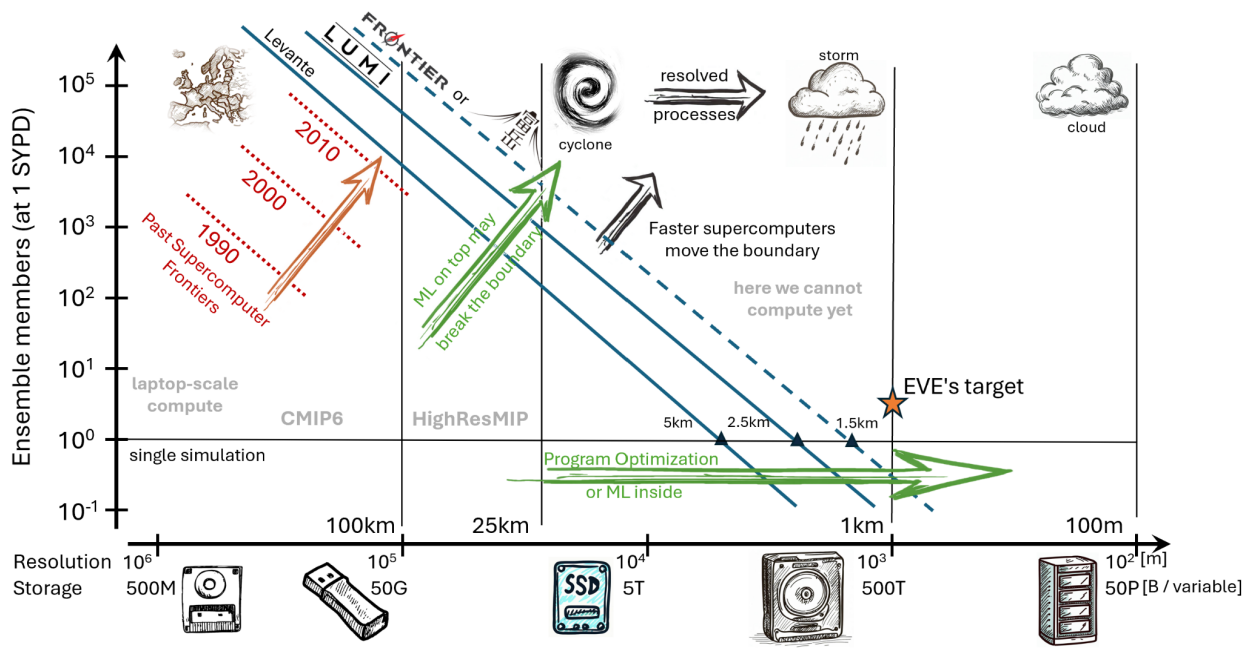

Figure 2: Resolution vs. number of ensemble members and the computational boundaries to be pushed with EVE. The x-axis shows increasing resolution and the number of ensemble members is shown on the y-axis. The x-axis also shows the approximate simulation output data size for one variable over 30 years. A central speed metric is Simulated Years per Day (SYPD) — it is limited by how simulations perform on today's computers. The illustration shows the limits of existing climate simulations on several machines as blue lines: ICON running on Levante at DKRZ, LUMI, and projected on Fugaku and Frontier, which we estimate to perform within 15% of each other for this workload based on their HPCG performance. In red, we outline the historic frontiers based on the available compute capability in the past. We also outline some opportunities for "ML inside" and "ML on top" in green [bauer23]. Typical model resolution today is 100 km, higher-resolution models use 25-50 km [haarsma16].

Based on a rough extrapolation of benchmarking data [schulthess18], we expect that top-class supercomputers such as Frontier or, soon, Alps should be able to execute optimized decadal climate simulations at 1 km resolution if the machine was exclusively dedicated to the task for multiple weeks. Yet, the produced data volume is expected to be many tens of exabytes [schulthess18].

It's Really All About (Accessing and Interacting With) The Data!

From a technical perspective, a climate simulation produces an output array of values at each grid point at specific output intervals (e.g., 15 minutes). The potentially many ensemble members sample the probability distribution of each of those values. So a simulation's output can be identified by a multi-dimensional tensor: The horizontal $i$, $j$ (surface) dimensions of the grid, the vertical $k$ (atmospheric or ocean) dimension, the time dimension $t$, and the ensemble dimension $e$ for each variable of interest (e.g., temperature).

A single variable, for example, the surface temperature, stored as 8-bit value sampling in 15 minute intervals for a typical 30-year simulation requires 0.5 PiB of storage. Yet, the temperature at different height levels controls cloud formation, such that one would need to store more than 100x this value in the vertical dimension [schulthess18]. Because the climate is described by many variables (tens to hundreds), and tens of simulations are required to sample internal variability, naive output strategies can easily produce many exabytes to characterize a single climate change scenario.

Given the exabyte storage requirements, EVE must offer a storage backend that supports highly-optimized domain-specific compression [huang22,kloewer21]. Furthermore, EVE's access protocols must be extremely lightweight and efficient to enable fastest low-overhead access to the data. Most importantly, such a planetary, federated data serving ecosystem for climate data that is globally accessible simplifies rapid development of ML applications on top and avoids data copies. In fact, simple access through a global index and compatible access protocols will foster collaboration, reproducibility, and progress.

We envision a simple interface that can be accessed by users with very different levels of sophistication (Fig. 3). EVE's data layer has different access roles: scientists, consultants and policymakers, and citizens and laypeople - each of which have different requirements. Experts want access to all data fields and can deal with vast unfiltered gridded variable data; consultants want interpreted access to the simulation data; and citizens want analyses of interest in a digestible format such as maps of risk of flooding or wildfires.

The raw data can be accessed through the Cartesian Data Interface as a sparse multi-dimensional tensor. The dimensions include the spatial dimensions i, j, and k of the grid, and additional ones to index the variable of interest, generating model and model configuration. The tensor is sparse to also support observation data as well as fill-in non-existing information. The latter could be offered using automated (e.g., ML-based) interpolation of non-existing elements and return an uncertainty for each queried point. This results in a query interface similar to getPoint(i, j, k, t, e, var, model, setup, …) that returns a point and its certainty. Similarly, region-queries return sub-tensors or neighborhoods. It could be served through RESTful APIs like the ECMWF's polytope server or a faster low-level cluster API using Remote Direct Memory Access (RDMA), for example. This could be invisible to users if EVE offers language bindings. A discovery interface would allow to enumerate dimensions such as all available simulation setups. Unifying this storage of climate information would require agreement on the same grid, e.g., HEALPix.

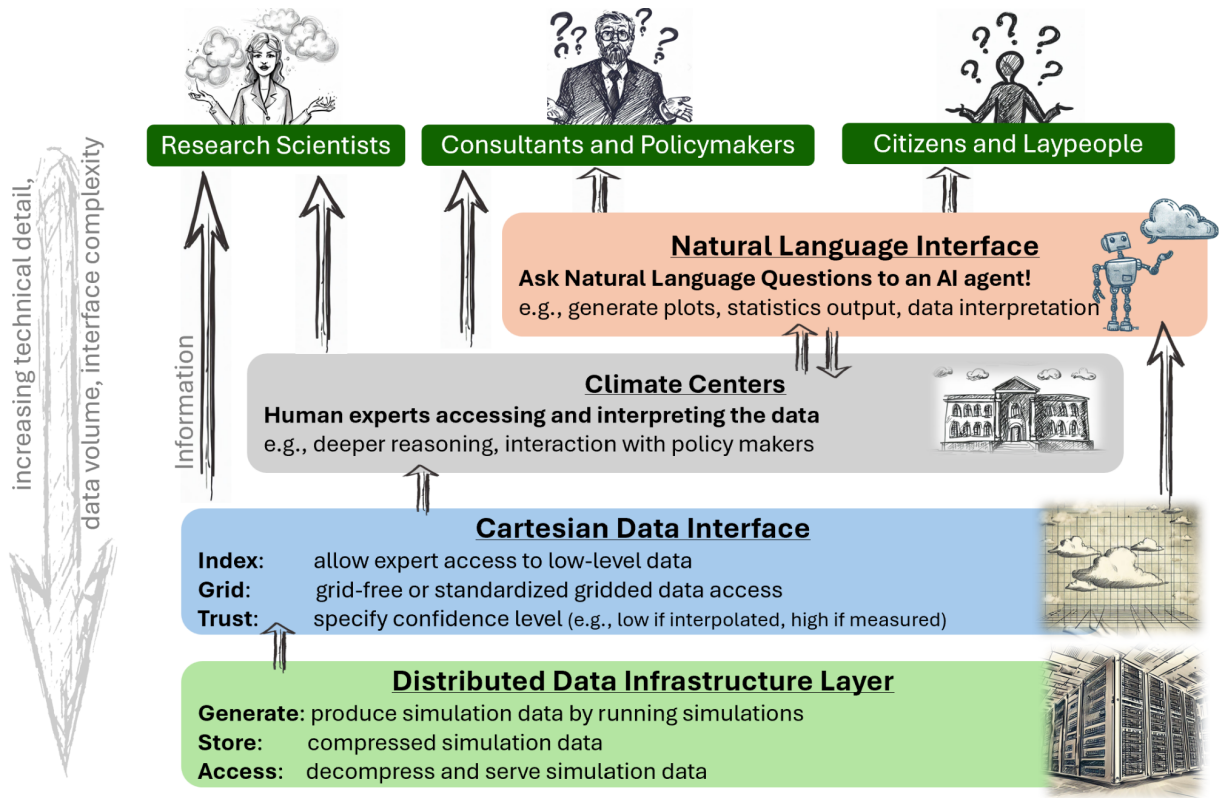

Figure 3: Information flow in a potential EVE data layer design

*Developing the data layer for EVE will be challenging but with exciting opportunities for innovation, especially for the application of ML.* For example, instead of compressing the data, checkpoints at specific time intervals as proposed by SimFS [girolamo19] could be used for resimulations (Figure 4) or ML-generated trajectories that faithfully reproduce the distribution of simulations between checkpoints (Fig. 5). Alternatively, innovative ML-based compression techniques could overfit model data to reproduce a tensor of data from its coordinates [huang22]. Furthermore, users could pre-register analyses for a large simulation run that will be executed on the fly, similar to beam-lines in CERN [schaer21].

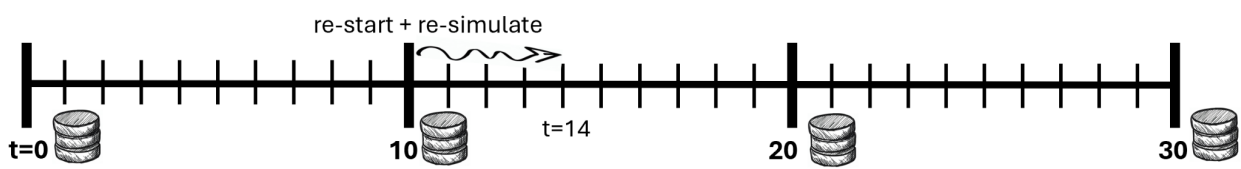

Figure 4: illustration of SimFS resimulation. We show 30 output steps in 15 minute intervals, each consists of many more simulation time-steps (not shown). During the simulation, data is only written at output steps divisible by 10, thus providing a 10x compression ratio. When, for example, step 14 is required, the simulation would be re-started at step 10 and run for 4 output steps forward [girolamo19].

## Initial Ideas for Using Machine Learning Techniques

Three main avenues for using ML techniques in the context of EVE are possible: (1) improving the simulation, (2) post-processing and storing the data, and (3) interpreting the data.

Ideas for improving the simulation itself abound. One approach is to replace parameterized parts of the simulation, such as radiation [chevallier00], with faster and potentially more accurate ML-based models. Another, more extreme, idea is to rely purely on data-driven forecasting and use ML-based weather models such as Pangu-Weather or FourCastNet. Yet, those may diverge quickly from reality if they are not constrained by physical laws, especially when used to extend into future climate states. An intermediate approach would be to use a SimFS-style checkpointing method running a simulation forward and storing checkpoints every k steps. Then, one would train a "tethered" model to converge to each of those checkpoints like a tethered boat or cable ferry. At the same time, the model could capture the statistics of multiple trajectories between the checkpoints (Fig. 5). Of course, the tethering points themselves should follow ensemble members from the overall distribution. The benefits are twofold, the trained ML-tethering model generates trajectories 100–1000 times faster than a traditional climate simulation, whilst simultaneously offering a 100–1000x compression on data storage requirements.

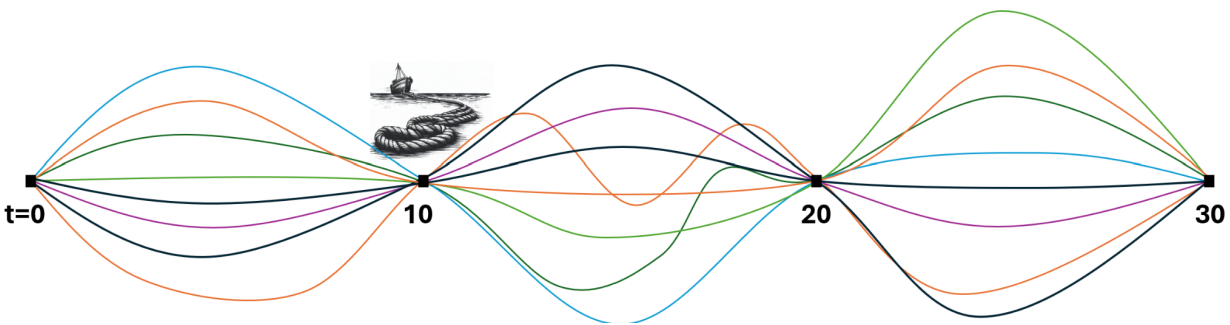

Figure 5: Eight colored trajectories are tethered at four points at 10-output steps intervals. This way, the overall simulation's drift is controlled while representing the statistics. ML methods such as generative AI could increase the number of steps between intervals to 100–1000.

A second avenue for ML use is post-processing to optimize data storage. One way to compress data would be to explicitly model the uncertainty in the ensemble dimension using simple statistics or advanced ML models [groenquist21]. Another opportunity could be to use ML models to compress the fields in the tensor directly by overfitting a model to reconstruct a block of the tensor data as exactly as possible. Earlier work has demonstrated compression rates of 1000x and more using this technique [dnn-comp].

The most important and probably biggest avenue for using ML techniques is interpreting the data, i.e., extracting actionable information [bauer23]. At the highest level, an interactive "interpretative agent" large language model would be able to answer questions about climate scenarios in natural language, potentially supported by output plots and data series. Fig. 3 shows this as Natural Language Interface, which could build on other ML models that extract information in a hierarchical way at lower levels. An exciting opportunity here is that ML not only helps achieve interactivity but also massively scales the extraction of actionable information in myriad user-defined, bespoke ways– exactly what is needed to realize EVE's vision. This area is largely unexplored but offers significant opportunities.

For each of those tasks, one could think about training foundation models that represent the latent distribution of climate data. One could store the inherent properties of climate data in an ML-derived latent space that could lead to faster compression using fine-tuning. A second foundation model could be the knowledge base for the interpretative agent language model.

## Conclusions and Outlook

The Earth Virtualization Engines (EVEs) stand as a critical intersection between computer, computational and climate sciences, necessitating collaboration among these sectors, industry, and academia. The technical challenges EVE poses are enormous - from creating high-resolution simulations to making Exabyte-level data accessible for all. However, these hurdles represent remarkable opportunities for innovation and cross-disciplinary collaboration. Together, we can make EVE a reality, not just a vision.

Acknowledgments

We thank all participants of the EVE summit in Berlin for many engaging discussions! TH was supported by the ADIA lab. We thank Lorena Barba for her editorial comments improving the quality of the manuscript.

References

[klinker00] Klinker, Ernst, et al. "The ECMWF operational implementation of four‑dimensional variational assimilation. III: Experimental results and diagnostics with operational configuration." Quarterly Journal of the Royal Meteorological Society 126.564 (2000): 1191-1215. https://doi.org/10.1002/qj.49712656417

[hersbach20] Hersbach, Hans, et al. "The ERA5 global reanalysis." Quarterly Journal of the Royal Meteorological Society 146.730 (2020): 1999-2049. https://doi.org/10.1002/qj.3803

[schneider17] Schneider, Tapio, et al. "Climate goals and computing the future of clouds." Nature Climate Change 7.1 (2017): 3-5. https://doi.org/10.1038/nclimate3190

[bauer23] Bauer, Peter, et al. "Deep learning and a changing economy in weather and climate prediction." Nature Reviews Earth & Environment (2023): 1-3. https://doi.org/10.1038/s43017-023-00468-z

[haarsma16] Haarsma, Reindert J., et al. "High resolution model intercomparison project (HighResMIP v1. 0) for CMIP6." Geoscientific Model Development 9.11 (2016): 4185-4208. https://doi.org/10.5194/gmd-9-4185-2016

[schulthess18] Schulthess, Thomas C., et al. "Reflecting on the goal and baseline for exascale computing: a roadmap based on weather and climate simulations." Computing in Science & Engineering 21.1 (2018): 30-41. https://doi.org/10.1109/MCSE.2018.2888788

[girolamo19] Di Girolamo, Salvatore, et al. "SimFS: a simulation data virtualizing file system interface." 2019 IEEE International Parallel and Distributed Processing Symposium (IPDPS). IEEE, 2019. https://doi.org/10.1109/IPDPS.2019.00071

[huang22] Huang, Langwen, and Torsten Hoefler. "Compressing multidimensional weather and climate data into neural networks." arXiv preprint arXiv:2210.12538 (2022).

[schaer21] Schär, Christoph, et al. "Kilometer-scale climate models: Prospects and challenges." Bulletin of the American Meteorological Society 101.5 (2020): E567-E587. https://doi.org/10.1175/BAMS-D-18-0167.1

[groenquist21] Grönquist, Peter, et al. "Deep learning for post-processing ensemble weather forecasts." Philosophical Transactions of the Royal Society A 379.2194 (2021): 20200092. https://doi.org/10.1098/rsta.2020.0092

[kloewer21] Klöwer, Milan, et al. "Compressing atmospheric data into its real information content." Nature Computational Science 1.11 (2021): 713-724. https://doi.org/10.1038/s43588-021-00156-2

[chevallier00] Chevallier, F., et al. "Use of a neural‐network‐based long‐wave radiative‐transfer scheme in the ECMWF atmospheric model." Quarterly Journal of the Royal Meteorological Society 126.563 (2000): 761-776. https://doi.org/10.1002/qj.49712656318

[prein15] Prein, Andreas F., et al. "A review on regional convection‐permitting climate modeling: Demonstrations, prospects, and challenges." Reviews of geophysics 53.2 (2015): 323-361. https://doi.org/10.1002/2014RG000475

[lorenz72] Lorenz, Edward. "Predictability: Does the flap of a butterfly's wing in Brazil set off a tornado in Texas?." (1972). https://mathsciencehistory.com/wp-content/uploads/2020/03/132_kap6_lorenz_artikel_the_butterfly_effect.pdf

[forster21] Forster, Piers, et al. "The Earth's energy budget, climate feedbacks, and climate sensitivity." The Physical Science Basis. Contribution of Working Group I to the Sixth Assessment Report of the Intergovernmental Panel on Climate Change. (2021). https://doi.org/10.1017/9781009157896.009

## Authors

Torsten Hoefler is a professor of computer science at ETH Zurich in Switzerland. His research interests include large-scale AI and HPC systems and applications in the area of large language models and climate sciences. Hoefler received his PhD in computer science from Indiana University. Contact him at htor@ethz.ch.

Bjorn Stevens is the managing director of the Max Planck Institute for Meteorology in Hamburg, Germany, and a professor at the University of Hamburg. He is broadly interested in the physics of climate. Stevens received his PhD in Atmospheric Science at Colorado State University. Contact him at bjorn.stevens@mpimet.mpg.de.

Andreas F. Prein is a researcher at the National Center for Atmospheric Research in Boulder, CO, USA. His research focuses on high-resolution modeling of mesoscale processes in the climate system including extreme weather events. Prein received his PhD in physics from the University of Graz. Contact him at prein@ucar.edu.

Johanna Baehr is a professor at the Institute of Oceanography at Universität Hamburg, where she leads a working group on 'Climate Modelling' which focuses on subseasonal to decadal climate prediction. She is the spokesperson of the Cluster of Excellence 'Climate, Climatic Change, and Society' CLICCS). Baehr received her PhD in Physical Oceanography from Universität Hamburg. Contact her at johanna.baehr@uni-hamburg.de.

Thomas Schulthess, is Professor of Computational Physics at ETH Zurich and the Director of the Swiss National Supercomputing Center (CSCS) in Lugano. Further to developing and operating a leading research infrastructures for extreme-scale computing and data, CSCS has been designing and operating the NWP systems for MeteoSwiss. Schulthess received his PhD in natural science from ETH Zurich and can be reached at schulthess@cscs.ch.

Thomas Stocker is Professor of Climate and Environmental Physics in the Physics Institute at the University of Bern. His interest is in the dynamics of the climate system over the past 1 million years and its future evolution, particularly anthropogenic climate change. He served as Co-Chair of the Intergovernmental Panel on Climate Change IPCC from 2008 to 2015. Stocker received his PhD in natural sciences from ETH Zürich. Contact him at stocker@climate.unibe.ch.

Professor John Taylor is currently Research Group Leader in CSIRO Data61, and Chief Computational Scientist at the Defence Science and Technology Group and Honorary Professor, College of Engineering & Computer Science at the Australian National University. My research interests include AI for science, computational and simulation science, climate change, global biogeochemical cycles, air quality and environmental policy, from the local to the global scale, spanning science, impacts and environmental policy. Contact him at John.Taylor@data61.csiro.au.

Daniel Klocke is the leader of the computational infrastructure and model development group at the Max Planck Institute for Meteorology. He is particularly interested in the development of the next generation of climate models with km-scale resolution and their application on large HPC systems. Klocke received his PhD in Meteorology from the University of Hamburg. Contact him at daniel.klocke@mpimet.mpg.de.

Pekka Manninen is the Director of Science and Technology at Advanced Computing Facility at CSC, the Finnish IT center for science. He has a long experience in supercomputing and supercomputing infrastructures and has been leading and involved in several pan-European e-infrastructure initiatives. He holds a Ph.D. in theoretical physics and is an Adjunct Professor at the University of Helsinki. Contact pekka.manninen@csc.fi.

Piers Forster is a Professor of Physical Climate Change and Director of the Priestley International Centre for Climate at the University of Leeds. He studies the causes and impacts of climate change. Piers holds a Ph.D. in meteorology from the University of Reading. Contact him at P.M.Forster@leeds.ac.uk.

Tobias Kölling is a Postdoc in the Max Planck Institute for Meteorology in Hamburg, Germany. His research interests include remote sensing, climate physics and attached data systems. Kölling received a Doctor of Natural Sciences from Ludwig-Maximilians-Universität in Munich. Contact him at tobias.koelling@mpimet.mpg.de.

Nicolas Gruber is Professor of Environmental Physics at ETH Zürich, Switzerland. He is interested in studying the Earth's biogeochemical cycles, especially that of carbon, and uses models and observations to determine how these cycles interact with the physical climate system. He received his Ph.D. in Environmental Physics from the University of Bern. Contact him at nicolas.gruber@env.ethz.ch.

Hartwig Anzt, is the Director of the Innovative Computing Lab (ICL) and Professor in the Electrical Engineering and Computer Science Department of the University of Tennessee. He also holds a Senior Research Scientist Position at Steinbuch Centre for Computing at the Karlsruhe Institute of Technology where he previously held a Junior Professorship in the Faculty of Computer Science. Hartwig Anzt holds a PhD in applied mathematics and specializes in numerical methods for the next generation hardware architectures. Contact him at hanzt@icl.utk.edu.

Claudia Frauen is a research software engineer at the German Climate Computing Center (DKRZ) in Hamburg, Germany. She is interested in HPC for weather and climate with a focus on strategies for performance portability to enable km-scale climate simulations. Claudia Frauen received her PhD in climate science from Christian-Albrechts-University Kiel and the GEOMAR Helmholtz Center for Ocean Research Kiel. Contact her at frauen@dkrz.de.

Florian Ziemen is a research software engineer at the German Climate Computing Center (DKRZ) in Hamburg, Germany. His interests include all things related to analysis and visualization of climate model output on HPC systems. Ziemen received his PhD in climate science from University of Hamburg. Contact him at ziemen@dkrz.de.

Milan Klöwer is a postdoctoral associate at MIT. His research interests include climate model development, low-precision computing and climate data compression. Klöwer received his PhD in climate computing from the University of Oxford. Contact him at milank@mit.edu.

Karthik Kashinath is a principal engineer and scientist at NVIDIA Corporation in the USA and co-leads NVIDIA's Earth-2 Initiative. His research interests include large-scale AI and HPC, and physics-informed machine learning. Kashinath received his PhD in aerospace engineering from the University of Cambridge. Contact him at kkashinath@nvidia.com.

Christoph Schär is Professor at the Institute for Atmospheric and Climate Science at ETH Zürich. His group is developing and exploiting high-resolution regional climate models on HPC platforms. He has a PhD in atmospheric dynamics from ETH Zürich. Contact him at schaer@env.ethz.ch.

Oliver Fuhrer, is Head of Numerical Prediction at MeteoSwiss, the national weather service of Switzerland. His research interests include high-resolution numerical weather prediction and HPC for weather and climate. Fuhrer received his PhD in physics from ETH Zurich. Contact him at oliver.fuhrer@meteoswiss.ch.

Bryan Lawrence is the University of Reading Professor of Weather and Climate Computing and a senior scientist in the UK National Centre for Atmospheric Science, NCAS. His research interests span atmospheric dynamics, climate science, and modeling and data technologies. Bryan received his PhD in atmospheric physics from the University of Canterbury in NZ. Contact him at bryan.lawrence@ncas.ac.uk.